\documentclass[letterpaper,11pt]{article}
\usepackage{tabularx} 
\usepackage{amsmath}  
\usepackage{graphicx} 
\usepackage[margin=1in,letterpaper]{geometry} 
\usepackage{revsymb}

\usepackage[final]{hyperref} 
\hypersetup{
	colorlinks=true,       
	linkcolor=blue,        
	citecolor=blue,        
	filecolor=magenta,     
	urlcolor=blue         
}
\usepackage{blindtext}

\usepackage{graphicx}  
\usepackage{subfigure}
\usepackage{multirow}

\usepackage{fancyhdr}
\usepackage{longtable}
\usepackage{parskip}
\usepackage[T1]{fontenc}
\usepackage{dcolumn}   

\usepackage{bm}        
\usepackage{amsfonts}  
\usepackage{amsmath}   
\usepackage{amssymb}   

\usepackage{mathrsfs}  
\usepackage{amsmath,amsfonts}
\usepackage{tcolorbox}
\usepackage{titlesec}

\usepackage{url}

\usepackage{graphicx}  
\usepackage{subfigure}
\usepackage{multirow}
\usepackage{authblk}

\usepackage{fancyhdr}
\usepackage{longtable}
\usepackage{parskip}
\usepackage[T1]{fontenc}
\usepackage{dcolumn}   
\usepackage{braket}

\usepackage{bbm}
\usepackage{tikz}
\usepackage{caption}
\usepackage{hyperref}
\usepackage{collectbox}
\usepackage{titlesec}
\usepackage{url}
\usetikzlibrary{positioning}

\newcommand{\qbinom}[2]{\Bigl[\,\begin{matrix}#1\\#2\end{matrix}\,\Bigr]_q}

\date{} 

\begin{document}

\title{Original $\mathbb F_1$ in emergent spacetime}
\author{Seyed Khaki\footnote{\href{https://afundas.com/}{Academy of Fundamental Studies}, Munich 80637, Germany\\ Email: \href{mailto:s.khaki@afundas.com}{s.khaki@afundas.com}}}

\maketitle
\setlength{\parindent}{3ex}

\begin{abstract}
The existence of a quantum field theory over the “field with one element” was first addressed in 2012 by Bejleri and Marcolli, where it was shown that wonderful compactifications of the graph configuration spaces that appear in the calculation of Feynman integrals, as well as the moduli spaces of curves, admit a $\mathbb F_1$ structure. Recently, we also examined some advantages of studying finite fields $\mathbb F_q$, wherein $\mathbb F_1$ represents the fundamental string with the Planck length which plays a fundamental role in the model. Such a role was briefly described, by comparing the similarity between the collapse of the spacetime concept probing the scales below the Planck length and the mathematical collapse of the 'field' concept at $q=1$. In this letter, we elaborate more on this role by explaining how Kapranov and Smirnov's perspective based on the Iwasawa theory is a perfect mathematical fit for the string theory. Particularly, their work suggests that the existence of $\mathbb F_1$ alone is sufficient to create other fields $\mathbb F_q$ emergent as its extensions, which exactly reflects the postulate of the string theory (where various vibrational modes of the fundamental string manifest as other fields). As support, a couple of evidence are provided that illustrate the physical importance of the Weyl group (known to be a reductive group over $\mathbb F_1$), and explain why the calculation of the amplitudes in the "amplitudes=combinatorial geometry" program (initiated by Arkani-Hamed et al. in 2013) exhibits a combinatorial nature and simplifies the exponentially growing ($\mathcal{O}(4^n)$) calculations of the Feynmann diagrams to a polynomially growing order ($\mathcal{O}(n^2)$) in the kinematic space of scattering data for the scalar theory with cubic interactions.
\end{abstract}

\section{Introduction}
Inspired by \cite{schnetz2009quantum}, the existence of a quantum field theory (QFT) over $\mathbb F_1$, the “field with one element”, was first studied in 2012 by Bejleri and Marcolli \cite{bejleri2013quantum}, where they explored the potential presence of an additional $\mathbb F_1$ structure for particular classes of algebraic varieties which innately emerge within the perturbative QFT. In particular, they showed that wonderful compactifications of the graph configuration spaces, that appear in the calculation of Feynman integrals in position space, as well as the moduli spaces of curves, admit a $\mathbb F_1$ structure. Unfortunately, such a bright idea was not adequately tracked. The second outstanding research avenue addressing a likely connection of $\mathbb F_1$ with physics is the studies of endomotive, presented in \cite{CONNES2007761} as a means of representing the formation of quantum statistical mechanical systems (QSMS) tied to number theory through arithmetic data, which began with the introduction of Bost-Connes (BC) system in \cite{bost1995hecke}. Remarkably, Connes, Consani, and Marcolli announced in \cite{connes2009fun} that the algebra and the endomotive of the QSMS of BC innately emerge by extension of scalars from $\mathbb F_1$ to rational numbers. Further, in \cite{chang2014quantum}, a Galois field quantum mechanics whose $q = 1$ limit becomes the classical mechanics is assessed.

On the other hand, by a different approach, we recently examined some advantages of studying finite fields $\mathbb F_q$ in \cite{khaki2023examination}, where at $q=1$ limit, the string length $\ell_s = q\times \ell_{Planck}$ becomes the Planck length representing the fundamental string, which, as its name suggests, plays a 'fundamental role' in the model. Such a role was briefly described by comparing the similarity between the collapse of the spacetime concept probing the scales below the Planck length and the mathematical collapse of the 'field' concept at $q=1$. In this letter, we elaborate more on this role in the next part, by proposing that Kapranov and Smirnov's perspective \cite{Kapranov1995} based on the Iwasawa theory is a promising mathematical fit for the string theory. Afterward, we present a brief tailored introduction to $\mathbb F_1$ to ensure the necessary grasp of its 'simplification' and 'original' role. In the last section, we provide a couple of evidence that illustrates the physical importance of the Weyl group (known to be reductive group over $\mathbb F_1$) and explain quantitatively why the calculation of the amplitudes in the "amplitudes=combinatorial geometry" program (initiated by Arkani-Hamed et al. in 2013 \cite{arkani2014amplituhedron}) exhibits a combinatorial nature and simplifies the exponentially growing calculations of the Feynmann diagrams to a polynomially growing order in the kinematic space of the scattering data.
 
\subsection{Proposal}
Inspired by a strong analogy between number fields and curves over finite fields, in 1959, Iwasawa founded his $Z_p$-extensions theory \cite{zbMATH03145290}, which investigates the growth of arithmetic objects in towers of number fields. In 1995, based on the Iwasawa theory that adjoining the roots of unity is analogous to producing extensions of a base field, Kapranov and Smirnov \cite{Kapranov1995} formed the profound perspective that the set of the $n$-th roots can be introduced as $\mathbb{F}_{\hspace{-0.1em}{1}^{\raisebox{0.4ex}{\scalebox{0.7}{$\hspace{0.1em}n$}}}}$, the field extensions of $\mathbb F_1$. In this manner, $\mathbb F_q$ is an algebra of dimension $\frac{q-1}{n}$ over $\mathbb F_{{1}^{\raisebox{0.4ex}{\scalebox{0.7}{$n$}}}}$, implying that all results of linear algebra over $\mathbb F_q$ must be reproduced by replacing the equivalent extension of the field with one element $\mathbb{F}_{\hspace{-0.1em}{1}^{\raisebox{0.4ex}{\scalebox{0.7}{$\hspace{0.1em}q-1$}}}}$. This suggests that only the existence of $\mathbb F_1$ is sufficient to create other fields emergent as its extensions, which exactly reflects the postulate of the string theory where various vibrational modes of the fundamental string manifest as other fields/particles. Thus, \textit{we propose that Kapranov and Smirnov's mathematical development is a perfect match for the string theory, where only one original field exists and its extensions manifest as other fields}.
\subsection{Original field}
In this part, we briefly introduce $\mathbb F_1$ emphasizing its 'original' and 'simplification' role (for a general overview see \cite{lorscheid2018f}). Although based on the standard definition of the ’field’, $\mathbb F_2$ is the smallest field, and a field with only one element does not exist, mathematicians had an idea of the meaning of projective geometry over that field. One could recognize $\mathbb F_1$ as a field-like algebraic object that behaves like a finite field with characteristic one if that field could exist. A more precise definition is that $\mathbb F_1$ is the multiplicative monoid \{0,1\} without the additive structure which addresses the $\mathbb F_1$ geometry as "non-additive geometry" (see \cite{pena2011mapping} for an overview of several geometries over $\mathbb F_1$). Studying geometry over $\mathbb F_1$ indeed uncovers the minimum amount of information (i.e., the most fundamental piece of data) about an object that enables examination of its geometric characteristics. This is also evident in the discovery of $\mathbb F_1$ in 1956 which Tits uncovered in \cite{tits1956analogues} (and Steinberg in \cite{steinberg1951geometric}), where he observed that projective geometries over $\mathbb F_q$ contain a meaningful analog at the minimal $q = 1$ limit. As an instance, over $\mathbb F_1$, the symmetric group $S_n$ on $n$ elements behaves as if it were a linear group,
\begin{equation}
\label{eq:1}
    GL_n(\mathbb F_q)  \ \ \ \begin{tikzpicture}[baseline={(0,-0.8ex)}]
    \draw [->, thick] (0.5,0) -- node[below] {$q\to1$} (2.5,0);
\end{tikzpicture}\ \ \ S_n
\end{equation} 
Here, the $q=1$ limit unveils the combinatorial skeleton/core of the Chevalley group $GL_n(\mathbb F_q)$. So, over $\mathbb F_1$, vector spaces (the objects on which $GL_n(\mathbb F_q)$ acts) reduce/descend to pointed sets (the objects on which $S_n$ acts) which one may imagine as \textbf{origin}. In this sense, $\mathbb F_1$ is the 'original' base (i.e., a terminal object lying below the integers) for an "absolute point" Spec$\:\mathbb F_1$ over which any algebraic scheme would sit. Such an original role in Tit's perspective includes many schemes of finite type over the integers that can be characterized by a polynomial with integer coefficients, serving as a counting function that counts $\mathbb F_q$ points of the scheme for all prime power values of $q$. Instances contain affine spaces, projective spaces, and Grassmannians.
\begin{equation}
\label{eq:2}
    \# \mathbb A^n(\mathbb F_q)=q^n  \quad\quad\quad \# \mathbb P^{n-1}(\mathbb F_q)=[n]_q  \quad\quad\quad \#Gr(k,n)(\mathbb F_q)= \qbinom nk
\end{equation}  
where $[n]_q=q^{n-1}+...+q+1$ is the Gauss number, $[n]_q!=\prod_{i=1}^{n}[i]_q$ is the Gauss factorial, and  $\ \qbinom nk =\frac{[n]_q!}{[k]_q![n-k]_q!}$ is the Gauss binomial. At $q=1$, the number of points will be minimized to $\mathbb F_1$-rational points
\begin{equation}
\label{eq:3}
    \# \mathbb A^n(\mathbb F_1)=1  \quad\quad\quad \# \mathbb P^{n-1}(\mathbb F_1)=n  \quad\quad\quad \#Gr(k,n)(\mathbb F_1)=\binom{n}{k}
\end{equation}
Now, let us view an example directly sourced from \cite{lorscheid2018f} to comprehend how $\mathbb F_1$ simplifies/reduces the information to its minimal meaningful level.

\begin{figure}[ht]
\centering
\begin{tikzpicture}[node distance = 8cm]
   \node (origin) {};
   \foreach \a in {1,...,14}{\draw (\a*360/7+90/7: 2cm) node [draw,circle,inner sep=2pt] (l\a) {};}
   \foreach \a in {1,...,7}{
      \draw (\a*360/7+270/7: 2cm) node [draw,circle,inner sep=1.8pt,fill=black] (p\a) {};
      \draw [-] (p\a) -- (l\a);
      \pgfmathtruncatemacro{\next}{\a + 1}
      \draw [-] (p\a) -- (l\next);
      \pgfmathtruncatemacro{\next}{\a + 2}
      \draw [-] (p\a) -- (l\next);
   }
   \foreach \a in {1,...,6}{\node[right=of origin] (\a) at (\a*360/3+30: 1.2cm) [draw,circle,inner sep=2pt] (l\a) {};}
   \foreach \a in {1,...,3}{
      \node[right=of origin] (\a) at (\a*360/3+90: 1.2cm) [draw,circle,inner sep=1.8pt,fill=black] (p\a) {};
      \draw [-] (p\a) -- (l\a);
      \pgfmathtruncatemacro{\next}{\a + 1}
      \draw [-] (p\a) -- (l\next);
      \pgfmathtruncatemacro{\next}{\a + 2}
    
   }
   \draw [->,thick] (3.5,0) -- node[below] {$q\to1$} (5.5,0) ;
\end{tikzpicture}
\caption{The geometry of $\mathrm{GL_3}(\mathbb{F}_q)$ for $q=2$ and its $q=1$ limit}
\label{figure:1}
\end{figure}
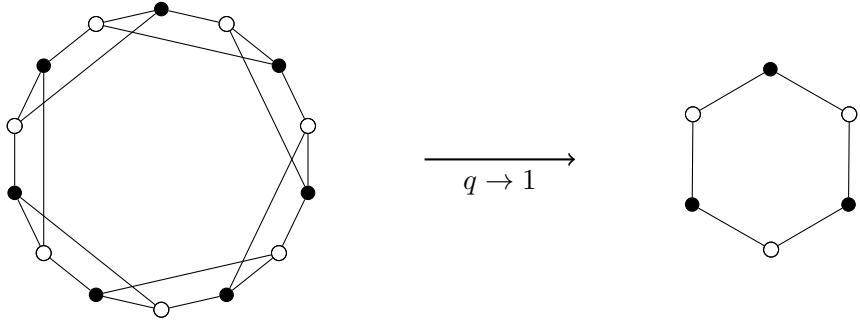

\textbf{Example} We illustrate the ideas of Tits in the example of $GL_3(F_q)$. In this case, we consider the two Grassmannians $Gr(1,F_q^3)$ and $Gr(2,F_q^3)$, whose points corresponds to the points and lines in $P^2(F_q)$, respectively. The incidence relation consists of pairs of a point $P$ and a line $L$ such that $P\in L$. In the case $q=2$, we calculate $\#Gr(1,F_2^3) = \qbinom 31 =  7$ and $\#Gr(2,F_2^3) = \qbinom 32 = 7.$ Moreover, note that every line contains $q+1=3$ points and that every point is contained in $q+1=3$ lines. The corresponding geometry is illustrated on the left hand side of Figure \ref{figure:1}, where the dots correspond to the points in $P^2(F_q)$, the circles correspond to the lines in $P^2(F_q)$, and an edge between a dot and a circle indicates that the corresponding point is contained in the corresponding line. Considering the limit $q\to 1$ yields the geometry for $S_3$ that consists of the sets $\Sigma(1,3)=\bigl\{\{1\},\{2\},\{3\}\bigr\}$ and $\Sigma(2,3)=\bigl\{\{1,2\},\{1,3\},\{2,3\}\bigr\}$. Note that every point of this geometry, i.e., an one element subset of $\{1,2,3\}$, is contained in $q+1=2$ lines, i.e., a $2$-subset of $\{1,2,3\}$. Similarly, every line contains $q+1=2$ points. The corresponding geometry is depicted on the right hand side of Figure  \ref{figure:1}.

\section{Supports and implications}
\subsection{Weyl group}
In his bold paper \cite{tits1956analogues}, Tits extends the analogy between $S_n$ and $GL_n(\mathbb F_q)$ to an analogy between the $\mathbb F_q$-rational points of a Chevalley group scheme $G$ and its Weyl group $W$ such that $ W = G(\mathbb F_1)$. That is, $\mathbb F_1$ unveils the Weyl group which is the core or essence of the main group. He indicates that, in the $q=1$ limit, the finite geometry attached to $G(\mathbb F_q)$ will be the geometry of the Coxeter group $W$. In this way, $\lim_{q\to 1} |G(\mathbb F_q)|/(q - 1)^{n_w}$ should count the number of elements of the Weyl group (where $n_w$ zeros of $G(\mathbb F_q)$ at $q=1$ are removed). E.g., for the exceptional Chevalley group $E_8(\mathbb F_q)$ we have $|E_8(\mathbb F_q)|=q^{120}({q}^{30} - 1) ({q}^{24} - 1) ({q}^{20} - 1) ({q}^{18} - 1)({q}^{14} - 1) ({q}^{12} - 1) ({q}^{8} - 1)({q}^{2} - 1)$, and therefore, $|W(E_8(\mathbb F_q))|=\lim_{q\to 1} \frac{|E_8(\mathbb F_q)|}{(q - 1)^{20}}=30\times24\times20\times18\times14\times12\times8\times2= 696\:729\:600$. This is the order we considered in \cite{khaki2023examination} to calculate the energy scale corresponding to the $E_8$, which here reveals its interpretation over $\mathbb F_1$. Over $\mathbb F_2$, the order blow up to $|E_8(\mathbb F_2)|\approx 3.3\times 10^{74}$. Hence,
\begin{equation}
\label{eq:4}
    |E_8(\mathbb F_2)|\approx 3.3\times 10^{74} \ \ \ \begin{tikzpicture}[baseline={(0,-0.8ex)}]
    \draw [->, thick] (0.5,0) -- node[below] {$q\to1$} (2.5,0);
\end{tikzpicture}\ \ \ |E_8(\mathbb F_1)|\approx 6.9\times 10^{8} 
\end{equation}
Now, let us cite the physical importance of the Weyl group. As established in \cite{cieciura1987physical} (and followed in \cite{cieciura1987weyl}), for a particle symmetry described by a Lie group $G$, not only the Weyl group $W(G)$ canonically acts in all zero weight spaces of $G$ (and therefore on observables), but also its action gives various physical relations among which are those inferred from the $G$-transformation properties of observables. Notably, it is shown that in the case of finite-dimensional representations, there exists a canonical action of the Weyl groups on observables such that many physical relations are a consequence of the Weyl group action rather than the action of the original group of root systems on observables. Thus, \textit{we claim that such physical importance originates from the fact that the Weyl group is a reductive group over $\mathbb F_1$}.

\subsection{"Amplitudes=Combinatorial Geometry" (ACG) program}
In 2013, Arkani-Hamed et al. started the ACG program, first by introducing the "amplituhedron" in \cite{arkani2014amplituhedron}, based merely on elementary combinatorial and geometric structures in the kinematic space of the scattering data that 'simplifies' the conventional calculation of the amplitudes to simpler counting problems without referencing to bulk spacetimes and Hilbert space. In the simplest theory of colored scalar particles - the Tr$\phi^3$ theory - considered in the ACG program, the $n$-point tree amplitude is typically a sum over $C_{n-2}=\frac{1}{n-1}\binom{2n-4}{n-2}$ number of Feynman diagrams (where $C_{n}$ is the $n$\textsuperscript{th} Catalan number) which is of the size $\mathcal{O}(4^n)$($cf$. \cite{arkani2023Counting} Outlook section, and \cite{arkani2023Multiplicity} Eq. 1.1). On the other hand, their approach contains curve integrals of an action built from $\mathcal{O}(n^2)$ piecewise linear functions, originating from the kinematic space for $n$ massless momenta whose dimensionality is ($cf$. \cite{arkani2018scattering} Eq. 2.3)
\begin{equation}
\label{eq:5}
    \binom{n}{2}-n=\frac{n(n-3)}{2}
\end{equation}
where $\binom{n}{2}$ is the number of Lorentz invariant dot products of momenta $\ p_i\ . \ p_j$ and $-n$ comes from the momentum conservation constraints. In this context, therefore, the 'simplification' means reducing the order $\mathcal{O}(4^n)$ of the Feynman diagrams method which grows exponentially with the number of particles to the order $\mathcal{O}(n^2)$ of the ACG method which grows as a polynomial with the number of particles ($cf$. \cite{arkani2023Counting} Outlook section). \textit{We claim that such a simplification is similar to the familiar reduction mentioned above in} Eq. \eqref{eq:1}, Figure \ref{figure:1}, and Eq. \eqref{eq:4}. That is,
\[
\text{Feynman method: amplitudes ($\mathbb F_2$)}\ \ 
\begin{tikzpicture}[node distance = 8cm]
   \draw [->,thick] (3.5,0) -- node[above] {$q\to1$} (5.5,0) ; 
\end{tikzpicture}
\ \ \text{ACG method: amplitudes ($\mathbb F_1$)}
\]
Now, let us quantitatively assess this claim. Over $\mathbb F_q$, the dimensionality of kinematic space is the q-analog of Eq. \eqref{eq:5} (for q-analogs compare Eq. \eqref{eq:2} and \eqref{eq:3})
\begin{equation}
\label{eq:6}
    \qbinom n2 - [n]_q = \frac{[n]_q\times [n-1]_q}{[2]_q} - [n]_q
\end{equation}
This yields $\scalebox{1.05}{$\frac{(q^{n}-1)/(q-1)\times (q^{n}-1)/(q-1)}{1+q}$}-(q^{n}-1)/(q-1)$ which is $\mathcal{O}(q^{2n})$. Therefore, at $q=2$, it grows as $\mathcal{O}(4^n)$, and at $q=1$, it becomes Eq. \eqref{eq:5} which is $\mathcal{O}(n^2)$. Notably, while the exponential behavior shows up for all $q\geq 2$, the polynomial behavior emerges only at $q=1$ \footnote{This may have some implications in computer science}.

Furthermore, it is worth mentioning that in \cite{arkani2023hidden}, Arkani-Hamed et al. recently highlighted the unified stringy picture too as they generalized their results to the stringy Tr$\phi^3$ amplitudes and stated "There is a unique shift of the kinematic data that preserves the zeros, and this shift is precisely the one that unifies colored scalars, pions, and gluons into \textit{a single object}". They also remark "there is a beautiful reason for the universality of these zeros, which turn out to be deeply related to a surprising relation between these colored theories revealed upon understanding a unified “stringy” descriptions of all these amplitudes. These stringy generalizations inherit all the zeros and factorization patterns of the field theory amplitudes and in fact generalize them to infinite new families of zero/factorization patterns. They also allow us to see that amplitudes for colored scalars, pions, and gluons are all given by a single function, expanded about different points in the kinematic space." 

In sum, this letter concludes that not only the quantum gravity in the emergent spacetime should enjoy $\mathbb F_1$ and its geometry, but also a fundamental theory of nature should embody finite combinatorial properties emerging from the number theory.


\bibliographystyle{unsrt}
\bibliography{main.bib}
\end{document}